\documentclass[11pt]{article}
\usepackage[dvips]{graphicx,psfrag}

\def\beqa{\begin{eqnarray}}
\def\eeqa{\end{eqnarray}}
\def\papertitlepage{%\baselineskip 3.5ex 
\thispagestyle{empty}}
\def\Title#1{\vspace{0.5cm}\begin{center}
 {%\large
\bf #1} \end{center}
\vspace{-0.3cm}}
\def\Authors#1{\begin{center} {\large #1} \end{center}}
\def\Abstract{\vspace{0.3cm}\begin{center} 
{%\large\bf Abstract
}
           \end{center} %\parbigskip
}
\begin{document}

\papertitlepage
\vspace*{-1 cm}
%\hfill

\Title{
A QUANTUM ANALYSIS ON RECOMBINATION\\ 
OF D-BRANES AND\\
ITS IMPLICATIONS FOR AN INFLATION MODEL
}
\Authors{{   TAKESHI SATO
} \\
 \vskip 0.5ex
{\itshape  Institute of Physics, University of Tokyo, \\
3-8-1 Komaba, Meguro-ku, Tokyo 153-8902 Japan\\
Email:tsato@hep1.c.u-tokyo.ac.jp
} \\
}
%%%%%%%%%%%%%%%%%%%%%%%%%%%%%%%%%%%%%%%
\Abstract
%%%%%%%%%%%%%%%%%%%%%%%%%%%%%%%%%%%%%%

\vspace{-1.5 cm}

{\footnotesize  A quantum-mechanical technique is used within the framework of U(2) 
super-Yang-Mills theory to investigate processes
after recombination of two D-p-branes at one angle.   
Two types of initial conditions are considered, 
one of which with $p=4$ is a candidate of inflation mechanism.
It is shown that the branes' shapes come to have three extremes 
due to localization of tachyon condensation. 
``Pair-creations'' of open stings connecting 
the recombined branes is also observed.
The appearance of closed strings is also discussed;
the decaying branes are shown to 
radiate non-vanishing gravitational wave, which may be interpreted as 
evidence of closed string appearance.
A few speculations are also given on implications of the above phenomena 
for an inflation model.}

\vspace{1 cm} 

%\section{Introduction}
%\setcounter{equation}{0}

This work is based on my talk on July 31 in the conference ``String Phenomenology 2003''.
More detail is presented in ref.\cite{tsato} (to be revised at that time). 
%\cite{tsato}.

Recombination of D-branes at angles is a process 
increasing its value
from both phenomenological 
and theoretical point of view, related to the presence of
tachyonic modes which appear as modes of open strings
between the two D-branes\cite{angle}.  
In the brane inflation models,
these are some of the promising candidates 
of the mechanism of inflation\cite{halyo,oneangle,tye1,2angles}. %
%\cite{hirano1}%\cite{ky}
%\cite{oneangle}%\cite{int2}%\cite{hirano2}
%\cite{tye1}\cite{2angles}. %\cite{bro}\cite{br1}.
Intersecting brane models are also, 
among various string constructions of Standard Model,
one class of hopeful
candidates\cite{bl1,chiral,ur1},
in some of which it has been proposed that
the recombination process occurs as Higgs mechanism% 
\cite{chiral,bachas}. 
%although for this case  
%multiple angle cases are rather favored. 
On the other hand, this system 
and process can be regarded as a generalized 
setting of $D\bar{D}$ system and its annihilation process, 
which has been studied thoroughly\cite{sen1,dd1,sen2}. 
%\cite{dd2}\cite{dd3}%
%\cite{sen2}. %\cite{tye2}. 
In this way, this is one of the most important phenomena 
to explore at present in string theory.

The behavior of the system after the recombination, however, 
had been almost unexamined until recently.
Suppose one consider a recombination of two D-branes at one angle, 
wrapping around some cycles in a compact space.
One might expect that after the recombination, 
each brane would begin to take a ``shorter cut'' and then make 
a damped oscillation around some stable configuration of branes,
while radiating RR gauge and gravitational waves (and some others),
leading to the stable configuration.
However, the case was that only the shape of the recombined branes 
at the initial stage was merely inferred\cite{pol}, 
though the final  
configuration can be determined in some cases\cite{2angles}.

Recently, K. Hashimoto and Nagaoka\cite{hn}
proposed that a T-dual of super Yang-Mills theory (SYM)
can be an adequate framework
to describe the process, and in the case of two D-strings at one angle and via 
classical analysis, they presumed how their shapes develop 
at the very initial stage\footnote{Recently, the process 
is also analyzed via tachyon effective field theory in 
ref.\cite{wh} and very recently,
subsequent work of ref.\cite{hn} was done by K. Hashimoto and W. Taylor. 
(Refer to K. Hashimoto's talk
in {\it Strings 2003}.)}

The main purpose of this work is to investigate 
the process after
the recombination via SYM
``more rigorously'' and to reveal what happens in 
the process;
we will focus on the two points: 
time-evolution of the D-branes' shape,
and behavior (appearance) of fundamental strings.
The former can be of value in that 
we describes time-dependent, i.e. 
dynamical behavior of curved D-branes though for 
rather short time (which enables us to discuss
e.g. gravitational waves radiated by the branes).  
%The former might also be regarded as a first step toward the
%understanding of branes' shapes after recombination, 
%which are responsible for physical quantities 
%in intersecting brane models. 
The latter point is what happens in the process itself.   
What we mean by ``more rigorously'' is that 
we set concrete initial conditions and make a quantum analysis:
These are necessary steps in order to understand precise behavior of 
the system, because the 
system, its tachyon sector in particular, 
is essentially a collection of inverse harmonic oscillators,
and its behavior depends crucially on
the initial fluctuations of the system,
which can only be evaluated appropriately via a quantum analysis.
%but has not been done at least in  this setting.

The second purpose of this work is to discuss
implications of the above two points (phenomena),
for the inflation scenario of the setting that two D-4-branes are
approaching each other at one angle as in
ref.\cite{oneangle},\cite{tye1},
aiming at the understanding of the whole process of the scenario.

For the above two purposes,
we consider two types of initial conditions for  
D-p-branes for each $p\ge 1$. 
%As we will discuss later,
In order to make a quantum analysis,
we need the fact that all the modes have positive frequency-squareds at
$t=0$. So,
the first one (denoted as case (I)) is that 
two D-p-branes have been overlapping until $t=0$ but are put
intersected at one angle $\theta$ at the instant $t=0$. 
The second one (the case (II)) is that one of parallel two
D-p-branes were rotated 
by a small angle and are approaching each other very slowly.
The case (I) is one of the simplest conditions, rather easier 
to study the process itself.
The case (II) is more practical;
the case with $p=4$ is one of the hopeful setting for 
the brane inflation scenarios\cite{oneangle,tye1}.
%From a practical viewpoint, the case (I) may seem less realistic,
%but it may also be able to be regarded as some local or rough
%approximation to more complicated setting of D-branes.

%The outline of this paper is summarized as follows:

Our notation and set-ups are;
we set $l_{s}= 1 $ (and revive it when needed),
and $g_{s} \ll 1$, %very small,
and consider D-p-branes in a 10D
flat spacetime with 
coordinates $x_{a}$ ($a=0,..,9$). The world-volumes are parametrized by
$x_{0},x_{1},\cdots x_{p}$.
We compactify space dimensions $x_{p},\cdots,x_{9}$
on a (10-p)-torus with periods $L_{a}$ for $a=p,\cdots,9$.
We denote $x_{0}$ as $t$ and $x_{p}$ as $x$ below.
The simplest setting of the initial condition for each case is represented in Fig.1
\begin{center}
\includegraphics[width=9cm]{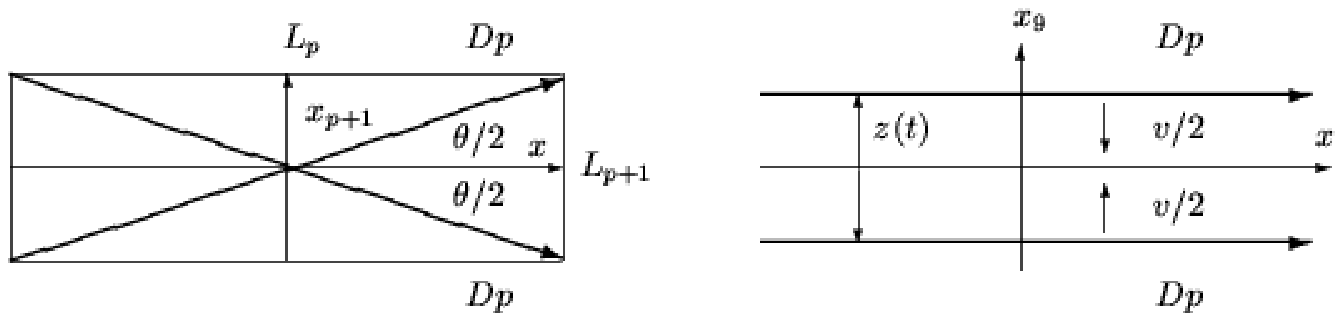}
%\caption{}

Figure 1: The initial configuration of D-p-branes ($z(t)=0$ for case (I))
\end{center}

The framework is a T-dual of SYM, 
a low energy effective theory of open strings around the D-branes
when distances between the branes are small, as in ref.\cite{hw,hn}.
We denote the U(2) gauge field by $A_{\mu}$ for
$\mu=0,1,..,p$ and adjoint scalar fields corresponding to
$x_{i}$, transverse to the branes,
as $X_{i}$ for $i=p+1,..,9$. 
We note that it holds
$L_{p}<\frac{1}{\theta}$ %\label{lcond1}
%\end{equation}
to satisfy the condition that
the displacements of branes are smaller than $l_{s}(=1)$.
The ``background'' D-p-branes for each case are
represented by\cite{hw}:
\begin{equation}
\begin{array}{ccc}
X_{p+1}= \left(
\begin{array}{cc}
\beta x & 0 \\
0 & -\beta x
\end{array}
\right),
 &
X_{9}= \left(
\begin{array}{cc}
z/2 & 0 \\
0 & -z/2
\end{array}
\right) ,
& X_{p+2}=\cdots =X_{8}=A_{\mu}=0.
\end{array}
\end{equation}
where $\beta\equiv \tan(\theta/2)$.
For the case (I) ($z(t)=0$),
this is T-dual
to the configuration of
two D-(p+1)-brane with a constant flux
$F_{p,p+1} = \beta$, in which
tachyonic modes %around the background 
were shown to appear in off-diagonal ones of
$A_{p}$ and $A_{p+1}$.\cite{baal,hw}.
For case (II) ($z(t)\ne 0$), 
since %$g_{s}$ and $\beta\equiv \tan(\theta/2)$
the force between the branes are so weak in this setting 
\cite{jabbari,pol,oneangle}, that
we approximate 
$z=z_{0} -v t$ for a constant $v$.
In both cases we can show that
potentially tachyonic modes appear
in the non-diagonal elements of $A_{p}$ and $X_{p+1}$.
So, we denote the fluctuations as
%including those of $A_{0}$ and $X_{9}$, as
\begin{equation}
\begin{array}{ccc}
A_{\hat{\mu}}= \left(
\begin{array}{cc}
0 & c_{\hat{\mu}}^{*} \\
c_{\hat{\mu}} & 0
\end{array}
\right), &
A_{p}= \left(
\begin{array}{cc}
0 & c^{*} \\
c & 0
\end{array}
\right)
, &
X_{p+1}= \left(
\begin{array}{cc}
0 & d^{*} \\
d & 0
\end{array}
\right)
\end{array}\label{ansatz2}
\end{equation}
where $\hat{\mu}=0,1,\cdots,p-1$. 

We evaluate time-evolution of typical values of 
the tachyonic fluctuations in the following; 
%%we assume zero or low temperature at $t=0$,
%we consider fluctuations aroud the D-branes and 
we define typical values of the fluctuations at time $t$ 
using VEV's of squares of the fluctuations, and 
make mode-expansion of them to diagonalize 
their second order terms in SYM action,
using $u_{k}(x_{\bar{\mu}})\equiv
e^{ik_{\bar{\mu}}x^{\bar{\mu}}}$ as for $x_{\bar{\mu}}$ 
($\bar{\mu}=1,\cdots,p-1$)
and $f_{n}\propto e^{-\beta x^{2}}H_{n}(\sqrt{2\beta}x)$ 
($H_{n}$ is Hermite function) as for $x$.\footnote{
$\frac{1}{\sqrt{\theta}} \ll L_{p}$ is needed here to use $f_{n}$.}   
Then, we can show that only the modes of $\tilde{c}(x_{\mu}) 
\equiv \frac{c-id}{\sqrt{2}}$ with
$n=0$ and the momentum-squared $k^{2} \le 2\beta$ are tachyonic ones.
So we obtain the typical value by using time-evolution of the wave function of 
$\tilde{c}_{0,k}$ and focusing only on the contribution of the tachyonic modes. 
We note that though we use second order approximation 
in the fluctuations, 
it is enough good until the tachyons blow up,
because higher order terms are suppressed by a factor 
$O(g_{s})$ since typical values of non-tachyonic modes are proportional to 
$1/\sqrt{T_{Dp}}=\sqrt{g_{s}}$. 
Defining $|c(x^{\mu})|_{{\rm typical}}^{2} =|d(x^{\mu})|_{{\rm typical}}^{2}
\equiv (A(t)f_{0})^{2}$,
we have for case (I), 
\begin{eqnarray} 
(A(t))^{2} 
=\frac{g_{s}\Omega_{p-2} }{2 (2\pi)^{p-1}}
\int_{0}^{\sqrt{2\beta}} k^{p-2}dk
[\frac{1}{k}+\frac{2\beta}{2\beta-k^{2}}
\sinh^{2}(\sqrt{2\beta-k^{2}}t)] \label{staticcfinal}
%\sqrt{\frac{2\beta}{\pi}}e^{-2\beta x^{2}} %\\
%\end{array}
\end{eqnarray}
 for D-p-brane for $p\ge 2$
\footnote{For D-strings, no integration with 
respect to $k$ is included, and $k^{2}$ is replaced by some 
regularization scale $(z^{(0)})^{2}$.},
where $\Omega_{p-2}$ is the volume of (p-2)-sphere and 
$f_{0}=\sqrt{2\beta/\pi}e^{-2\beta x^{2}}$.
For the case (II), the square bracket in (\ref{staticcfinal})
is replaced by a complicated function.  
We do not present it here (see ref.\cite{tsato}),
but the important thing is that we have obtained explicit functions 
of typical tachyonic fluctuations for each initial condition.
We note that in both cases, 
the tachyon condensation is localized around $x=0$ with 
the width $1/\sqrt{\beta}$ due to $f_{0}$.

The information of the shape is obtained from a  
transverse U(2) scalar field 
by diagonalizing  its VEV's (i.e. choosing a certain gauge)
and looking at its diagonal parts\cite{hn}.
(For case (II), a non-commutativity
between VEV's of scalar fields appears, implying their uncertainty
relation. 
We proceed with discussion by concentrating on a VEV of one field,
leaving alone another.)
Then, the formula for the shape of one of the recombined branes is
\begin{eqnarray}
y\equiv\sqrt{(\beta x)^{2} + |d|^{2}}=\sqrt{(\beta x)^{2}
+ \frac{A(t)^{2}}{2}\sqrt{\frac{2\beta}{\pi}}e^{-2\beta x^{2}} }.
\label{braneshape1}
\end{eqnarray}
This is the same form as obtained in ref.\cite{hn}.
However, we can make a precise analysis on the shape of the branes at
arbitrary $t$ based on (\ref{braneshape1}),
since we have the explicit function of $A(t)$ for each case
including the overall factor and fine coefficients;
%(in (\ref{staticcfinal}) or (\ref{sstring})); %or (\ref{movingcfinal}));
we can know explicitly
when and which applicable condition of the approximations
(SYM, WKB and some others) breaks.

Here, we show that
a seemingly queer behavior
occurs; the shape of each brane deviates from the approximate
hyperbola and comes to have three extremes as in Fig.2.
This happens for the cases of D-p-brane for $p\ge 2$,
and is also expected to happen for the case of D-strings, 
as we discuss below.\footnote{
The appearance of such a shape was already discussed,
but was {\it denied} in the case of D-strings in ref.\cite{hn}.}
\begin{center}
\includegraphics[width=6cm]{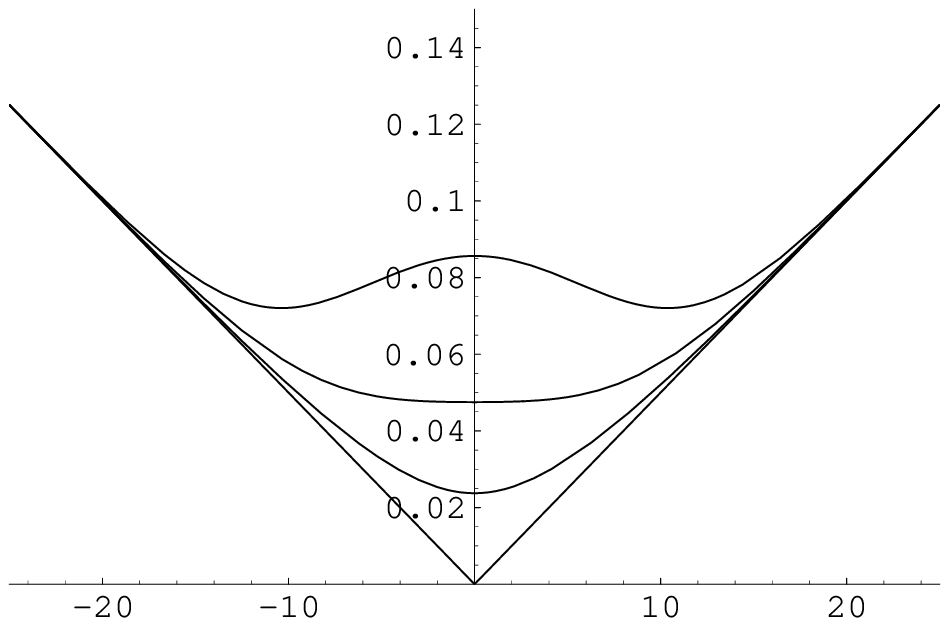}
%\caption{}

Figure 2: Time-evolution of the shape of a D-brane after recombination 
\end{center}
The values of $x$ giving extremes of (\ref{braneshape1})
are formally given by
$x=0,
%\pm\sqrt{\frac{1}{4\beta}\ln \frac{2 A^4}{\pi\beta}},
\pm\sqrt{ \{ \ln (2 A^4/\pi\beta) \} /4\beta }$,
which means that each  has three extremes
when
\begin{equation}
A>(\pi\beta/2)^{1/4}\equiv A_{{\rm critical}}.\label{multi1}
\end{equation}
The most important condition to be checked is
that of the WKB approxiamtion:
higher order terms of the fluctuations in the action
do not disturb the behavior.
%Since only $c$ and $d$ include the tachyonic field,
%we estimate the orders of their second order terms $S_{2}$ and
%fourth order terms $S_{4}$. 
%Then,
%the condition for $|S_{4}|\ll |S_{2}|$ is written as
The condition gives
\begin{equation}
A(t)\ll \sqrt{(2\pi)^{p-1/2}/
\Omega_{p-2}(2\beta)^{(p-2)/2}}.
\label{wkbdp}
\end{equation}
We can easily see that for a small $\theta\simeq 2\beta$ and $p\ge 2$,
the inequality (\ref{wkbdp}) is sufficiently
compatible with (\ref{multi1}).
The other conditions, especially that of SYM, that the displacement of
branes should be
smaller than $l_{s}$ ($A(t)< l_{s}$) and D-branes should have
some low velocity %($\dot{y}\ll 1(=c)$ at all the value of $x$),
are also shown easily to be satisfied.
Therefore, {\it the shape of the recombined D-p-brane for $p\ge 2$ surely
come to have three extremes}.
This happens essentially due to the localization of tachyons
around $x$,
since the factor $e^{-2\beta x^{2}}$ in (\ref{braneshape1}) is directly
responsible for the multiple extremes.
For the case of D-strings,
the case is a bit more subtle, but if we solve numerically the
equation of motion for $\tilde{c}_{0}(t)$
we can see that $|\tilde{c}_{0}(t)|$ really develop to exceed
$A_{{\rm critical}}$ in (\ref{multi1}) before any approximations break.

The physical interpretation of this seemingly queer behavior  is
as follows;
the energy released via tachyon condensation
pushes the recombined branes away from each other, but
it is given only to the local part of the branes around $x=0$
due to localization of tachyon condensation, so, only
the part is much accelerated.
Though the D-branes have a large tension,
they also have a large inertia, and
when the given energy of the local part is large enough,
the branes extend, surpassing the tension, to form three extremes.
That is, {\it localization of tachyon condensation or that of
the released energy, and the (large) inertia causes the shapes of branes
to have three extremes.}
We note that this is the ``physical'' gauge where the 
d.o.f. of each open string starting from and ending on certain branes
directly corresponds to the d.o.f. of each U(2) matrix element,
since only in this gauge each of the diagonal elements represents 
the position of each brane.

Next, we examine the behavior of open strings
based the typical value of the electric flux
since electric charges on
D-branes correspond to ends of open strings attached to
the D-branes\cite{cm}.
We note that we have to study the flux {\it in the physical gauge}.  
An important thing is that even after choosing the ``physical gauge''
(in which $X_{p+1}$ is diagonalized),
only the off-diagonal elements of the typical flux
$F_{0p}=\partial_{t} A_{p}$ blow up, since it holds
\begin{equation}
\begin{array}{cc}
F_{0p}|_{{\rm typical}}^{{\rm (phys)}}&
%=U^{-1}\frac{1}{\sqrt{2}}
%\left(
%\begin{array}{cc}
%0 &  |\partial_{t}\tilde{c}|_{{\rm typical}}
%e^{i \epsilon}\\
%|\partial_{t}\tilde{c}|_{{\rm typical}}
%e^{-i \epsilon} & 0
%\end{array}
%\right) U\nonumber\\
%& 
=\frac{1}{\sqrt{2}}\left(
\begin{array}{cc}
0 & |\partial_{t} \tilde{c}|_{{\rm typical}}
e^{i \epsilon'} \\
|\partial_{t}\tilde{c}|_{{\rm typical}}
e^{-i \epsilon'} & 0
\end{array}
\right)
\end{array}
\end{equation}
where $e^{i \epsilon'}$ is a
phase factor.
The blow-up of the off-diagonal elements in this gauge
means that fundamental strings connecting the two
D-branes are vastly created.\footnote{
This point was also discussed very recently by K. Hashimoto and W. Taylor
in Hashimoto's talk in``Strings 2003''.}
In addition, since the true VEV of (linear) $F_{0p}$ vanishes,
so does the total electric charge. That is,
these are effectively ``pair-creations''
of the strings.
Thus, we conclude that {\it vast number of open string pairs
connecting the two
D-branes are created after the recombination of D-branes.} 
(For estimation of their typical number, see ref.\cite{tsato}.)
We note that (at least at the early stage of the decay) about half 
the released energy is used to create the open string pairs, so
its energy turns into that of the open strings rather rapidly.

Furthermore, we discuss (expect) the behavior of the system, 
in particular fundamental strings, 
beyond our approximations and SYM.
If the distance between the D-branes becomes larger than $l_{s}$,
it is difficult to imagine that the created open strings extend
unlimitedly, because the string has its tension.
It also seems unnatural if one take into account increase of entropy.
Then, what will happen? Since open strings are created {\it in pairs}
with {\it no net NSNS gauge charges},
it is expected that {\it each pair of open strings connecting the branes
are cut to pieces to form some closed strings and two open strings},
each of the latter of which has its both ends on each of the two D-branes.
That is, such a picture seems to arise that the decaying (annihilating)
D-branes leave  vast number of closed strings
behind (and radiate some of them) while producing many open strings
which start from and end on each brane.
This may be regarded as a generalization of the picture for $D\bar{D}$
annihilation that only closed stings are left after the
annihilation\cite{dd1}.

Then, what evidence can we present to support the expectation ?
We can show  that the recombined D-branes surely radiate 
gravitational wave;
The angular distribution of the energy flux due to the radiation
$d^{2}E/dtd\Omega=\ \ <(\frac{d^{3}I_{ij}}{dt^{3}})^{2}
-2(n_{i}\frac{d^{3}I_{ij}}{dt^{3}})^{2}
+\frac{7}{8}(n_{i}n_{j}\frac{d^{3}I_{ij}}{dt^{3}})^{2}>$
in the direction $n_{i}$ 
is shown to be non-vanishing 
when $n_{i}$ is perpendicular to the $x_{p}x_{p+1}-$plane.
$I_{ij}$ is the mass quadrupole moment of the D-branes
\begin{eqnarray}
I_{ij}=T_{p}\sum_{A=1}^{2}\int d^{p}x\
\sqrt{1+(\partial_{x}x_{i}^{A})^{2}}
[x^{A}_{i}x^{A}_{j}-\frac{1}{9}\delta_{ij}(x^{A}_{k})^{2}]
\end{eqnarray}
where $A$ denotes each of the branes
and $x_{i}$ in $I_{ij}$
is to be evaluated at the retarded time.
Gravitational waves should be represented by closed strings
in string theory, so this may be regarded as evidence of
appearance of closed strings.
Since we know time-evolution behavior of the D-branes,
we can compute it within some approximations.
(See revised version of ref.\cite{tsato} more detail.)

Finally, we speculate on  implications of the above results for 
a brane inflation scenario.
Let us suppose that either of the following properties is (would be)
a generic one (though it might be a too optimistic extrapolation):(i)
Decaying D-branes continue to create vast number of open string pairs,
or
(ii) they would keep leaving vast number of closed strings behind
while producing open strings on each brane,
in both cases using some amount of the released energy.
Then, it might be expected  that the dissipation of the energy released
via tachyon condensation
would be rather large, i.e. damping factor would be large and
only a few times of, or no oscillations
would occur.
Thus, if this setting is applied to the inflation
scenario,
reheating might be efficient.
Furthermre, in the inflation model, (reheated) fermion and gauge fields
have to be produced,
but the mechanism how they are produced in this setting of the inflation
model has not been clarified.
We speculate about a scenario of the mechanism based on our analysis:
vast number of creations of
the open string pairs connecting the branes
and the open strings on each of the branes
might directly correspond to the fields.
It would be interesting to find a framework describing the process in
the region of distance other than SYM
or beyond SYM, and discuss the above possibilities.

%%%%%%%%%%%%%%%%%%%%%%%%%%%%%%%%%%%%%%%%%%%%

\section*{Acknowledgments}

I especially  thank Hidehiko Shimada for many fruitful
discussions, encouragement and much adequate advice.
I also thank to Prof. K. Hashimoto and Mr. Nagaoka for useful 
communication.
I also like to thank Dr. Taro Tani for many useful discussion 
and encouragement.
I would also like to thank Prof. H. Tye for helpful
comments. 
Finally, I would like to thank the organizers of
``String Phenomenology 2003'' for their hospitality.
This work is supported by JSPS under 
Post-doctorial Research Program (No.13-05035).

\end{document}